\documentclass[aps,pre,%
onecolumn,preprint,%
amssymb,amsmath,%
nopreprintnumbers,%
showpacs,%
showkeys,%
fleqn,%
eqsecnum,%
endfloats*,%
byrevtex]{revtex4}
\usepackage{calc}
\usepackage{dcolumn}
\usepackage[final]{graphicx}
\usepackage{color}
\usepackage{curves}
\begin{document}
\title{Simulation study of random sequential adsorption of mixtures
on a triangular lattice}
\author{I. Lon\v{c}arevi\'{c}}
\affiliation{Faculty of Engineering,
Trg D. Obradovi\'{c}a 6, Novi Sad 21000, Serbia}
\author{Lj. Budinski-Petkovi\'{c}}
\affiliation{Faculty of Engineering,
Trg D. Obradovi\'{c}a 6, Novi Sad 21000, Serbia}
\author{S. B. Vrhovac}
\email{vrhovac@phy.bg.ac.yu}
\homepage{http://www.phy.bg.ac.yu/~vrhovac/}
\affiliation{Institute of Physics,
P.O. Box 68, Zemun 11080, Belgrade, Serbia}
\date{\today}
\begin{abstract}
Random sequential adsorption of binary mixtures of extended objects on a
two-dimensional triangular lattice is studied numerically by means of Monte
Carlo simulations. The depositing objects are formed by self-avoiding random
walks on the lattice. We concentrate here on the influence of the symmetry
properties of the shapes on the kinetics of the deposition processes in
two-component mixtures. Approach to the jamming limit in the case of mixtures
is found to be exponential, of the form: $\theta(t) \sim
\theta_{jam}-\Delta\theta\; \exp (-t/\sigma),$ and the values of the parameter
$\sigma$ are determined by the order of symmetry of the less symmetric object
in the mixture. Depending on the local geometry of the objects making the
mixture, jamming coverage of a mixture can be either greater than both
single-component jamming coverages or it can be in between these values.
Results of the simulations for various fractional concentrations of the objects
in the mixture are also presented.
\end{abstract}
\maketitle
\section{Introduction}
\label{sec:intro}
Random sequential adsorption (RSA) has attracted considerable interest due to
its importance in many physical, chemical, and biological processes. In two
dimensions (2D), RSA is a typical model for irreversible and sequential
deposition of macromolecules at solid/liquid interfaces. Some examples of the
wide range of applicability of this model include adhesion of colloidal
particles, as well as adsorption of proteins to solid surfaces, with relaxation
times much longer than the formation time of the deposit. A comprehensive
survey on RSA and cooperative sequential adsorptions is given by Evans
\cite{Evans93}.  Recent surveys include Privman \cite{Privman00,CadilheAP07},
Senger \emph{et al.} \cite{SengerVS00} and Talbot \emph{et al.}
\cite{TalbotTVTVV00}.

The simplest RSA model is defined by the following three rules: (i) objects are
placed one after another in random position on the substrate; (ii) adsorbed
objects do not overlap; and (iii) adsorbed objects are permanently fixed at
their spatial positions (neither diffusion nor desorption from the surface are
allowed). When the surface is saturated by adsorbed objects so that no further
objects can be placed, the system reaches the jamming limit.  The RSA models
are broadly classified into continuum models and lattice models on the basis of
the nature of the substrate. A quantity of central interest is the coverage
$\theta(t)$, which is the fraction of the total substrate area occupied by the
adsorbed objects at time $t$. Asymptotic approach of the coverage fraction
$\theta(t)$ to its jamming limit $\theta_{jam} = \theta(t \to \infty)$ is known
to be given by an algebraic time dependence for continuum systems
\cite{Feder80,Swendsen81,Pomeau80}, and by exponential time dependence for
lattice models \cite{BarteltP90,NielabaPW90,BudinskiPKL97a,BudinskiPKL97}.

An important issue in RSA is the influence of the shape of the depositing
objects on the dynamics of irreversible deposition. RSA of many different
geometric objects has been studied. For instance, Khandkar \emph{et al.}
\cite{KhandkarLO00} have studied RSA of zero-area symmetric angled objects on a
continuum substrate for the full range ($0\,^{\circ} - 180\,^{\circ}$) of
values of the arm angle $\phi$ and have observed that $\theta_{jam} - \theta(t)
\sim t^\alpha $ as expected. The value of the exponent $\alpha$ exhibits a
crossover near $\phi = 0\,^{\circ}$ or $180\,^{\circ}$, and is significantly
lower in the case of the angled objects than in the case of needles.  Wang and
Pandey \cite{WangP96} have studied the kinetics and jamming coverage in  RSA of
self-avoiding walk chains on a square lattice.  They reported that the growth
of the coverage $\theta(t)$ to its jamming limit in the intermediate time
regime can be described by a power law $\theta(t) \sim \theta_{jam} -
c/t^\gamma$. In contrast to theoretical predictions for the RSA of polydisperse
object of regular shapes they find that effective exponent $\gamma$ depends on
the chain length.  They observed a crossover from a power-law variation of the
coverage fraction $\theta(t)$ in the intermediate time regime to an exponential
growth in the long time, especially for short chains. Budinski and Kozmidis
\cite{BudinskiPKL97a,BudinskiPKL97} have carried out the extensive simulations
of single-layer irreversible deposition using objects of different sizes and
rotational symmetries on a square and triangular lattice.  They fitted the
asymptotic approach of the coverage fraction $\theta(t)$ to its jamming limit
$\theta_{jam}$ by the exponential time dependence:
\begin{equation}
\theta(t) \sim \theta_{jam}-\Delta\theta\; \exp (-t/\sigma),
\label{eq:rhoIRSA}
\end{equation}
where $\Delta\theta$ and $\sigma$ are parameters that depend on the shape and
orientational freedom of depositing objects
\cite{BudinskiPKL97a,BudinskiPKL97}. The shapes with the symmetry axis of a
higher order have lower values of $\sigma$, i.e., they approach their jamming
limit more rapidly. The parameter $\Delta\theta$ decreases with the object size
for the same type of shape. Later, by studying a reversible RSA of extended
objects on a triangular lattice, Budinski \emph{et al.} \cite{BudinskiPPJV05}
showed that the growth of the coverage $\theta(t)$ above the jamming limit to
its steady-state value $\theta_{\infty}$ is described by a pattern $\theta(t)=
\theta_{\infty} - \Delta\theta E_\beta [-(t/\tau)^\beta]$, where $E_\beta$
denotes the Mittag--Leffler function of order $\beta\in(0,1)$.  The parameter
$\tau$ is found to decay with the desorption probability $P_-$ according to a
power law $\tau=A\;P_-^{-\gamma}$.  Exponent $\gamma$ is the same for all the
shapes, but parameter $A$ depends only on the order of symmetry axis of the
shape.  This confirms the crucial role of the geometrical character of the
objects in deposition dynamics.

In comparison to the irreversible deposition of pure depositing objects, very
little attention has been given to the RSA of two or more species of different
shape and/or size although the latter problem is inherent in many experimental
situations \cite{HarnarpSGK01,HarnarpSGK02}. The role of polydispersity has
been studied numerically in a wide variety of conditions. Examples include,
binary mixtures \cite{MeakinJ92,BonnierLP92}, mixtures of particles obeying a
uniform size distributions \cite{MeakinJ92}, mixtures with Gaussian size
distributions \cite{MeakinJ92,AdamczykSZW97}, power-law size distributions
\cite{BrilliantovAKK96}, etc.  Theoretical works were restricted only to a
binary mixtures of particles with very large size differences
\cite{TalbotS89,BarteltP91,TarjusT92}, power-law size distributions
\cite{Krapivsky92,BrilliantovAKK96}, or general continuous size distributions
\cite{TarjusT91}.  The motivation of our present work comes from Barker and
Grimson \cite{BarkerG88}, who investigated the adsorption of mixtures of
lattice objects of different shapes but of the same size on a square lattice by
means of computer simulation.

In this paper we study the irreversible deposition of two-component mixtures of
extended objects on a 2D triangular lattice by Monte Carlo simulations.
Simulations are performed for objects of various shapes. The depositing shapes
are modeled by directed self-avoiding walks on 2D triangular lattice.  A
self-avoiding shape of length $\ell$ is a sequence of \emph{distinct} vertices
$(\omega_0, \ldots, \omega_l)$ such that each vertex is a nearest neighbor of
its predecessor, i.e., a walk of length $\ell$ covers $\ell +1$ lattice sites.
Examples of such walks for $\ell=1, \ldots, 6$ are shown in
Table~\ref{tab:pure_objects}. On a triangular lattice objects with a symmetry
axis of first, second, third, and sixth order can be formed. Rotational
symmetry of order $n_s$, also called $n$-fold rotational symmetry, with respect
to a particular axis perpendicular to the triangular lattice, means that
rotation by an angle of $2\pi/n_s$ does not change the object. Here we focus
our interest on the influence of the order of symmetry axis of the shape on the
kinetics of the deposition processes in two-component mixtures.

The paper is organized as follows. Section \ref{sec:simulat} describes the
details of the simulations. We give the simulation results and discussions in
Sec.~\ref{sec:res}. Finally, Sec.~\ref{sec:concl} contains some additional
comments and final remarks.
\section{Definition of the model and the simulation method}
\label{sec:simulat}
In order to make a systematic approach to this problem and to be able to
compare the results for the mixtures to the results for the single-component
systems, the simulations were also performed for pure shapes shown in Table I.
The results are obtained for all the shapes that may have a different long-time
behavior of $\theta(t)$ and different jamming coverages $\theta_{jam}$ for the
lengths of the walks $\ell = 1,2$ and $3$. The number of different shapes
increases very fast with the length of the walk \cite{Jansen04} and for the
greater lengths we investigated a few characteristic objects for each length.
We made at least two representative objects for each order of symmetry.

At each Monte Carlo step  a lattice site is selected at random. If the selected
site is unoccupied, one of the six possible orientations  is chosen at random
and deposition of the object is tried in that direction. We fix the beginning
of the walk that makes the shape at the selected site and search whether all
successive $\ell$ sites are unoccupied. If so, we occupy these $\ell + 1$ sites
and place the object. If the attempt fails, a new site is selected, and so on.
After long enough time a jamming limit is reached when there is no more
possibility for a deposition event.

It should be noted that the results in Table I differ from the estimates given
in \cite{BudinskiPKL97}. In the present work we deal with the above described
conventional model of RSA, while in Ref. \cite{BudinskiPKL97} the end-on model
of RSA has been used. In the end-on model the depositing object always checks
all possible directions from the selected site. If the object cannot be placed
in any of the six orientations, the site is denoted as inaccessible. The
jamming limit for the end-on model is slightly larger than for the conventional
model and the approach to the jamming limit is faster.

The set of binary mixtures used in our simulations is shown in Table
\ref{tab:mixtures}.  Each mixture $(x)+(y)$ is composed of two lattice objects
$(A)$ -- $(U)$ from Table \ref{tab:pure_objects}.  In the case of mixtures, at
each Monte Carlo step  a lattice site is selected at random, one of the objects
that make the mixture is selected at random and deposition of the selected
object is tried in one of the six orientations.  If the attempt fails, a new
site  and a depositing object is selected at random. The jamming limit is
reached when neither of the objects can be placed in any position on the
lattice.

The Monte Carlo simulations are performed on a triangular lattice of size
$L=120$. Periodic boundary conditions are used in all directions. The time is
counted by the number of attempts to select a lattice site and scaled by the
total number of lattice sites. The data are averaged over 500 independent runs
for each shape and each mixture of shapes.
\section{Results and discussion}
\label{sec:res}
For all the objects from Table I and for all the mixtures from Table II plots
of \linebreak $\ln(\theta_{jam}-\theta(t))$  vs $t$ are straight lines for the
late stages of deposition. These results are in agreement with the exponential
approach to the jamming limit of the form (1.1), with parameters $\sigma$, $A$
and $\theta_{jam}$ that depend on the shape of the depositing object, i.e. on
the combination of the objects making the mixture.

Some of the plots for the pure objects are shown in Fig. 1. We can notice that
there are four groups of lines with different slopes, corresponding to shapes
with different order of symmetry.  The values of the parameter $\sigma$ are
determined from the slopes of the lines and they are given in Table I.
According to $\sigma$, all shapes can be divided into four groups:
\begin{itemize}
\item[a)] shapes with a symmetry axis of first order with $\sigma \simeq 5.9$;
\item[b)] shapes with a symmetry axis of second order with $\sigma \simeq 3.0$;
\item[c)] shapes with a symmetry axis of third order with $\sigma \simeq 2.0$;
\item[d)] shapes with a symmetry axis of sixth order with $\sigma \simeq 0.99$.
\end{itemize}
This means that the rapidity of the approach to the jamming limit depends on
the order of symmetry of the shape and the approach is slower for less
symmetric shapes. The symmetry properties of the shapes have a crucial
influence on the filling of small isolated targets on the lattice that are left
for deposition in the late times of the process. Namely, a shape with symmetry
axis of higher order has a greater number of possible orientations for
deposition into an isolated location and an enhanced probability for the
adsorption.

Kinetics of the irreversible deposition of mixtures is illustrated in Fig. 2
where the plots of $\ln(\theta_{jam}-\theta(t))$ vs $t$ are given for some
combinations of the shapes from Table I. These plots are straight lines for the
late times of deposition, suggesting that in the case of mixtures the approach
to the jamming limit is also exponential. In Fig. 2 lines with four different
slopes can be noticed. The value of the parameter $\sigma$ is determined by the
order of symmetry of the less symmetric object in the mixture and the values
are:
\begin{itemize}
\item[a)] $\sigma \simeq 11.6$ for the mixtures including an object with a
symmetry axis of first order;
\item[b)] $\sigma \simeq 5.65$ for the mixtures that contain at least one
object with symmetry axis of second order and no objects with symmetry axis of
first order;
\item[c)] $\sigma \simeq 3.97$ for the mixtures that contain at least one
object with symmetry axis of third order and no objects with symmetry axis of
lower order;
\item[d)] $\sigma \simeq 0.985$ for the mixtures of the objects with symmetry
axis of sixth order.
\end{itemize}
In the late stages of deposition  the less symmetric objects have to try all
possible orientations, so they are responsible for the approach to the jamming
limit. Deposition of mixtures is slower in comparison to the deposition of pure
objects, except for the objects with symmetry axis of sixth order for which the
parameter $\sigma$ has the same values for the pure shapes and for the mixture.
The retardation of the adsorption shows that each of the islands of the
connected unoccupied sites is specific to a particular shape. If we imagine an
island of unoccupied sites such that an adsorption of shape $(x)$ is possible,
prior to a succesful $(x)$-shape adsorption there will probably be some
rejected attempts of $(y)$-shape adsorption.  Therefore the overall process
will proceed more slowly than each of the individual adsorption processes.

Jamming configurations consist of clusters of objects of the same type. A few
typical jamming configurations are shown in Figs. 3 a) -- c).  These clusters
are more prominent for elongated and for symmetric objects. Namely, at very
early times of the process the depositing objects do not "feel" the presence of
the already deposited ones and are adsorbed randomly onto the surface.
However, in the late stages of deposition the depositing objects must fit into
small empty regions that favors the formation of clusters.  If one examines the
evolution of domains in the case of mixture $(S)+(U)$ (Fig. 3 c)), it can be
observed that growth of domains precipitated during early growth is
substantially frustrated in the case of angled object $(U)$. This is the cause
of the severe limitation on the size of domains in the case of angled objects
$(U)$ as compared to those in the case of $k$-mers $(S)$.

Value of $\theta_{jam}$ for a mixture depends on the local geometry of the
objects making the mixture. Values of the jamming coverage for different
combinations of depositing objects are given in Table II. Qualitatively, we
could say that it depends on the probability that the neighboring sites of an
adsorbed object would be blocked by another adsorbing object from the mixture.
For the same types of shapes the jamming coverage decreases when the size of
the objects increases.

For RSA of mixtures of $k$-mers of various lengths $\ell_1$ and $\ell_2$ on a
square lattice it was found \cite{MannaS91} that the jamming coverage for a
mixture $\theta_{jam}^{(\ell_1 + \ell_2)} $ is always greater than either of
the jamming coverages for the lengths $\ell_1$ and $\ell_2$. Comparing the
results from Table I and Table II we can see that for the mixtures of objects
of various shapes this is not always the case. For a number of combinations of
depositing objects given in Table II the jamming coverage for a mixture has
greater values than the jamming coverages for the pure shapes making the
mixture.  However, there are also mixtures such as $(E)+(F),\; (O)+(Q),\;
(O)+(R),\; (P)+(Q),\; (S)+(T),\; (S)+(U)$ that have a lower jamming coverage
than one of the components. The jamming coverage of the mixtures is still
greater than the  jamming coverage of the other component.  The mutual feature
of these mixtures is that the jamming coverages of their components differ
significantly.

We have also performed extensive simulations in order to investigate the
deposition processes for various fractional concentrations $r^{(x)}$ and
$r^{(y)}$ of the shapes $(x)$ and $(y)$ in the reservoir, i.e. for various
probabilities for choosing one of these shapes for a deposition attempt. As an
illustration, here we give the results for the jamming coverages and for the
parameter $\sigma$ for a few combinations of the objects and for various
fractional concentrations of depositing objects. In Figs. 4 a) -- c) jamming
coverages obtained for various compositions of mixtures $(x)+(y)=
(B)+(D),(S)+(T), (T)+(U)$ are shown vs the fractional concentration $r^{(x)}$
$(r^{(y)}=1-r^{(x)})$. We can see that $\theta_{jam}$ varies monotonously with
the fractional concentration of one of the objects, growing with the
concentration of the object with greater jamming coverage of the pure shapes.
Depending on the combination of the objects and on their fractional
concentrations  it can be either greater than both jamming coverages of the
pure shapes making the mixture or it can be lower than the higher
single-component jamming coverage, but still higher than the other. When there
is a large difference in the jamming coverages for the single-component
depositions, such as in the case of the mixture $(S)+(T)$ (Fig. 4 b)), the
jamming coverage for the mixture can be in between these coverages, especially
for the high fractional concentrations of the object with lower
single-component $\theta_{jam}$.

Dependence of the parameter $\sigma$ on the fractional concentration $r^{(x)}$
is shown in Fig. 5. The behavior of $\sigma$ differs from case to case. For
example, for the mixture $(B)+(D)$ (Fig. 5 a)), $\sigma$ has a minimal value
for fractional concentration $r^{(x)} \simeq 0.6$ and the process slows down
when one of the fractional concentrations grows at the expense of the other. On
the other hand, for the combination $(T)+(U)$, the value of $\sigma$ increases
with the fractional concentration of the more symmetric object $(T)$ (Fig. 5
c)). At first sight, this is somewhat surprising, since object $(T)$ has a
symmetry axis of sixth order and the lowest value of $\sigma$ in the case of a
single-component deposition, and object $(U)$ has a symmetry axis of first
order and the highest value of $\sigma$ in the case of a single-component
deposition. Indeed, the more symmetric objects reach their jamming coverage for
a short time, but there are still left some small empty regions where the
deposition of the less symmetric objects is possible. The lower is the
fractional concentration of the latter objects, the longer is the time
necessary to find all these empty regions left for the deposition.  Despite
these differences, the value of $\sigma$ has larger values for a mixture than
for the pure shapes for all fractional concentrations of the  components, so we
can generally say that the deposition process is always slower for a mixture
than for the pure shapes making the mixture.
\section{Concluding remarks}
\label{sec:concl}
Kinetics of irreversible deposition of mixtures on a triangular lattice has
been studied by Monte Carlo simulations. A systematic approach has been made by
examining a wide variety of shapes and their combinations. The approach to the
jamming limit was found to be exponential for all the shapes and all the
mixtures. The rapidity of the approach depends only on the symmetry properties
of depositing objects. For two-component mixtures the value of the parameter
$\sigma$ is determined by the order of symmetry of the less symmetric object in
the mixture.

It was shown that the kinetics of irreversible deposition fastens with the
increase of the order of symmetry of the shape.  On the contrary, the
adsorption-desorption process of asymmetric shapes is faster than the same
process of more symmetric shapes \cite{BudinskiPPJV05}. The reason for both
effects lies in the filling of small empty regions in the late stages of the
processes.  In the case of irreversible deposition, objects with symmetry axis
of higher order have greater number of possibilities for deposition into fixed
small empty regions in the late stages of deposition, while in the reversible
case the enhanced rate of single particle adsorption prolongates the approach
to the steady state coverage. Namely, in the reversible case, when the value
$\theta_{jam}$ is reached, the rare desorption events are generally followed by
immediate readsorption. These single-particle events do not change the total
number of particles. However, when one badly placed particle desorbs and two
particles adsorb in the opened location, the number of particles is increased
by one. These collective events are responsible for the density growth above
$\theta_{jam}$. The increase of the order of symmetry of the depositing object
enhances the rate of single-particle readsorption. This extends the mean
waiting time between consecutive two-particle events and the approach to the
steady state coverage is slower.

Jamming configurations for RSA of mixtures consist of clusters of blocked sites
and of clusters of objects of the same type. The dimensions of these clusters
are greater for elongated and for symmetric objects. Jamming coverage for a
mixture is always greater than the  jamming coverage of the component with
lower $\theta_{jam}$ and it is often greater than either of the jamming
coverages of the components making the mixture.

When the fractional concentration of one component grows at the expense of the
other component in the mixture, the jamming coverage varies monotonously. It
increases when the concentration of the object with greater single-component
jamming coverage increases. On the other hand, the rapidity of the approach to
the jamming coverage doesn't always show a monotonic behavior. For many
combinations there is a minimal value of $\sigma$, i.e. the jamming limit is
reached in the shortest time, for a certain value of fractional concentration
of one of the components making mixture.
\begin{acknowledgments}
This research was supported by the Ministry of Science of the Republic of
Serbia, under Grant No. 141035.
\end{acknowledgments}
 
\newpage
\textbf{Figures}
\newpage
\begin{center}
\includegraphics[width=\columnwidth, height=\columnwidth]{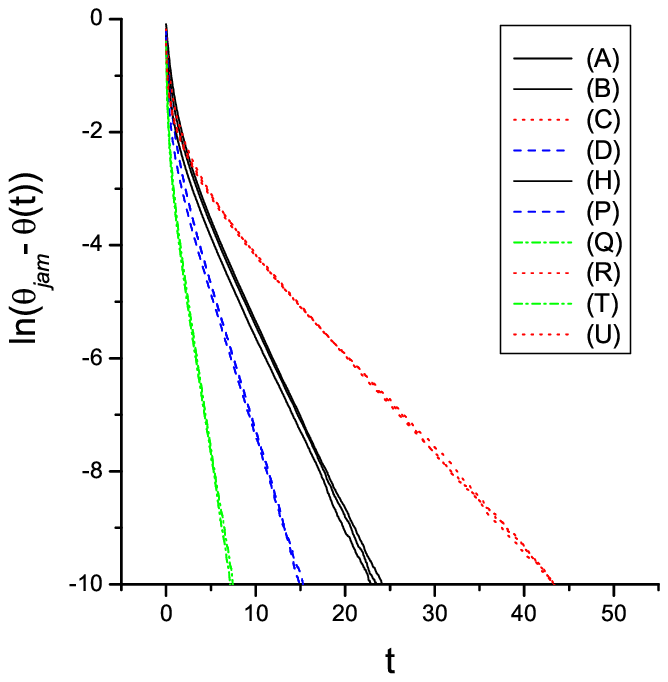}
Figure 1
\end{center}
(Color online) Plots of $\ln(\theta_{jam}-\theta(t))$ vs $t$ for various
objects from Table I.  Different slopes correspond to different symmetry
orders.
\newpage
\begin{center}
\includegraphics[width=\columnwidth, height=\columnwidth]{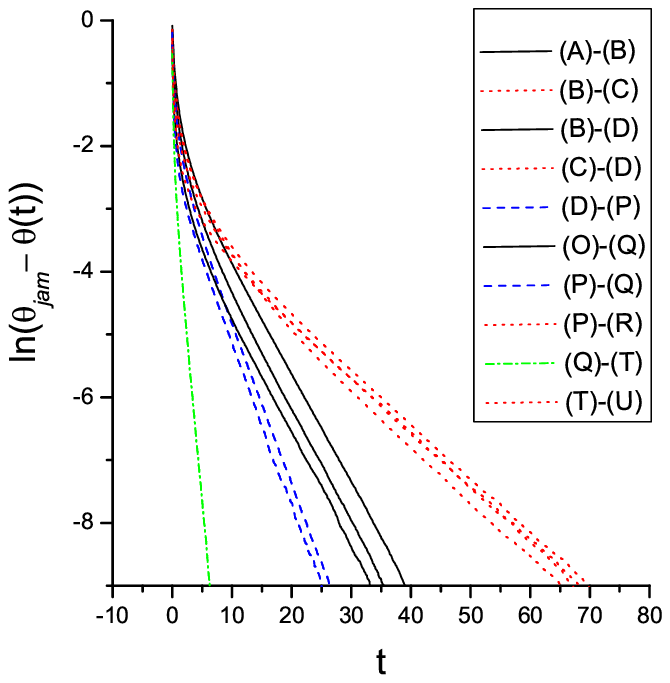}
Figure 2
\end{center}
(Color online) Plots of $\ln(\theta_{jam}-\theta(t))$ vs $t$ for various
mixtures from Table II. The slopes are determined by the order of symmetry of
the less symmetric object in the mixture.
\newpage
\begin{center}
\includegraphics[width=\columnwidth, height=\columnwidth]{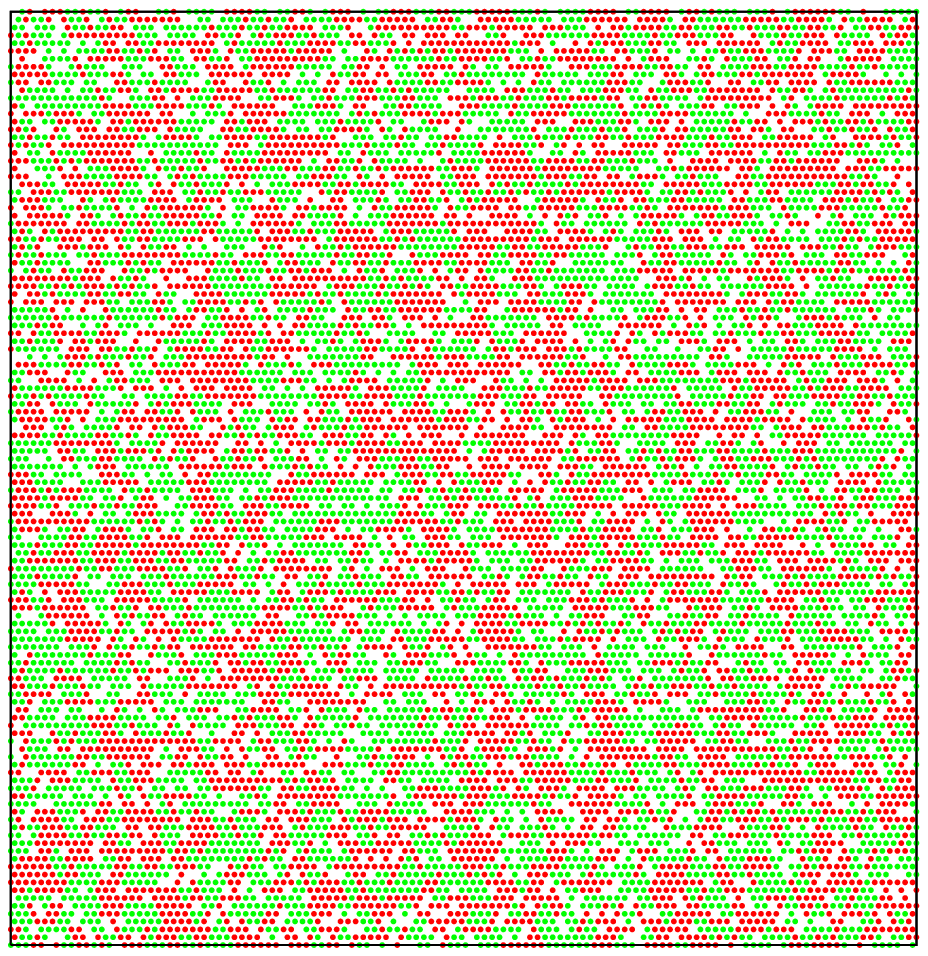}
Figure 3 a)
\end{center}
(Color online) Snapshots of patterns formed during the RSA of mixture $(B) +
(D)$ correspond to jamming state; $(B)$-red, $(D)$-green.
\newpage
\begin{center}
\includegraphics[width=\columnwidth, height=\columnwidth]{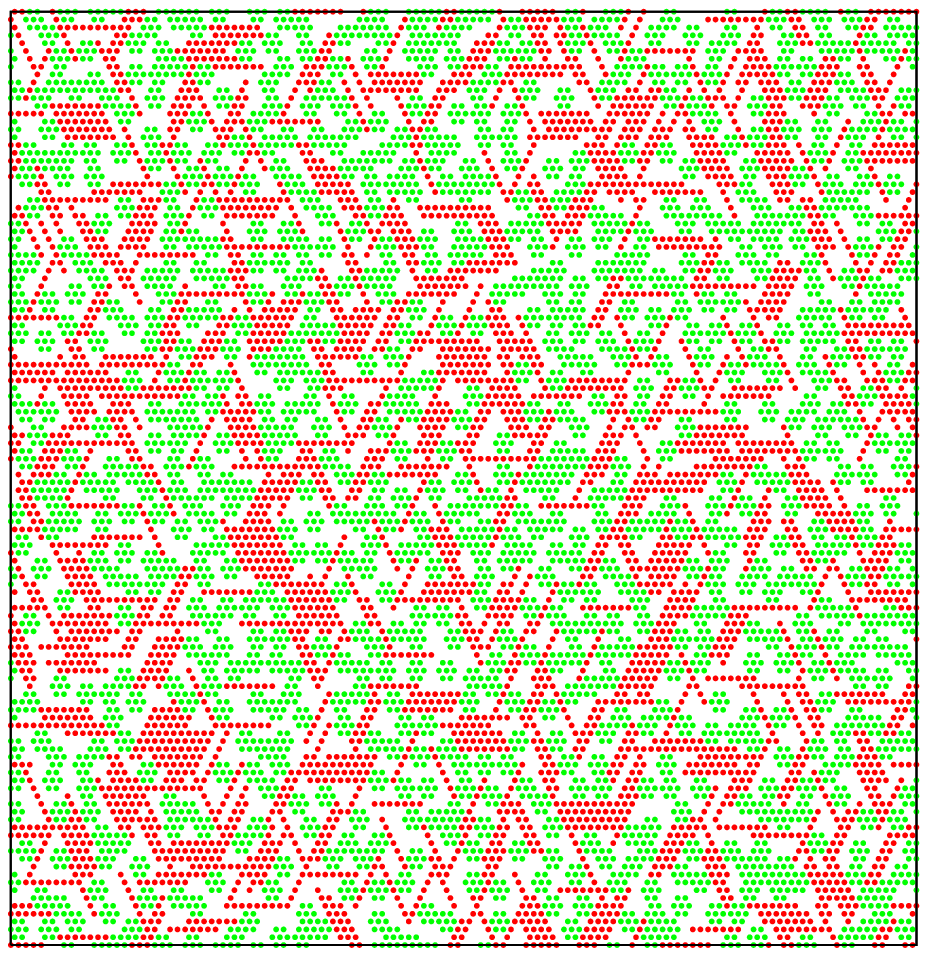}
Figure 3 b)
\end{center}
(Color online) Snapshots of patterns formed during the RSA of mixture $(S) +
(T)$ correspond to jamming state; $(S)$-red, $(T)$-green.
\newpage
\begin{center}
\includegraphics[width=\columnwidth, height=\columnwidth]{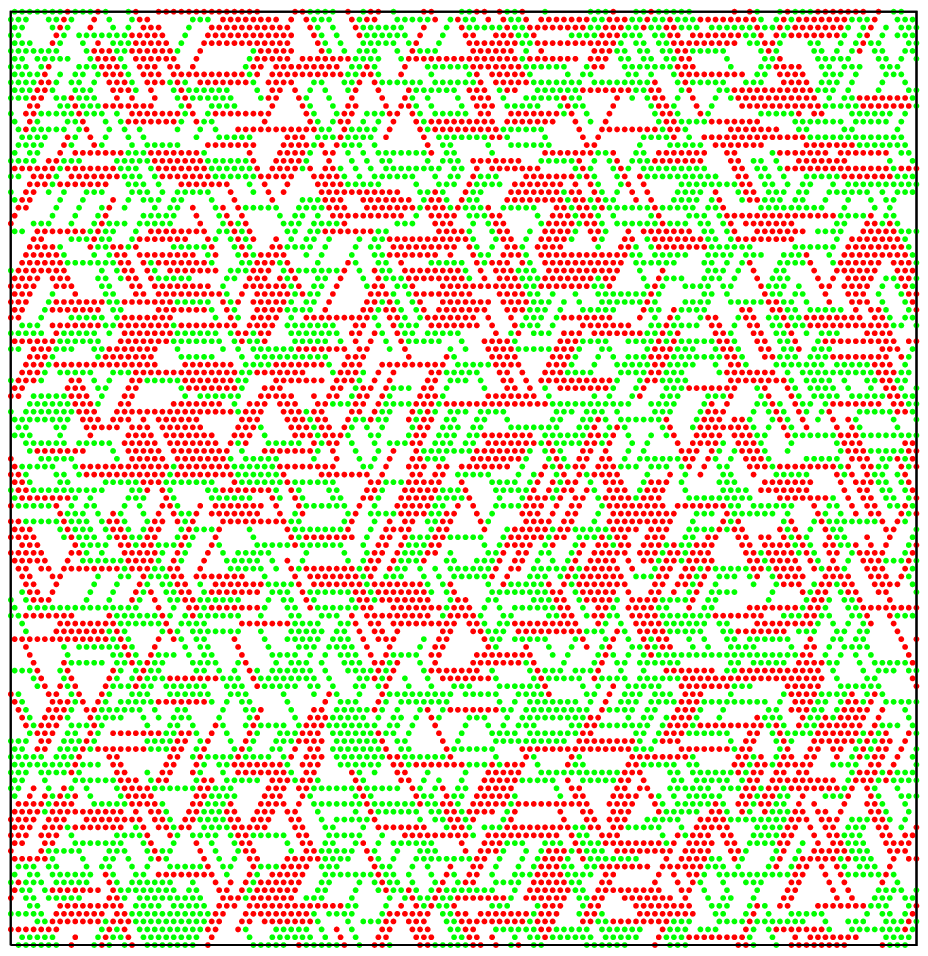}
Figure 3 c)
\end{center}
(Color online) Snapshots of patterns formed during the RSA of mixture $(S) +
(U)$ correspond to jamming state; $(S)$-red, $(U)$-green.
\newpage
\begin{center}
\includegraphics[width=\columnwidth, height=\columnwidth]{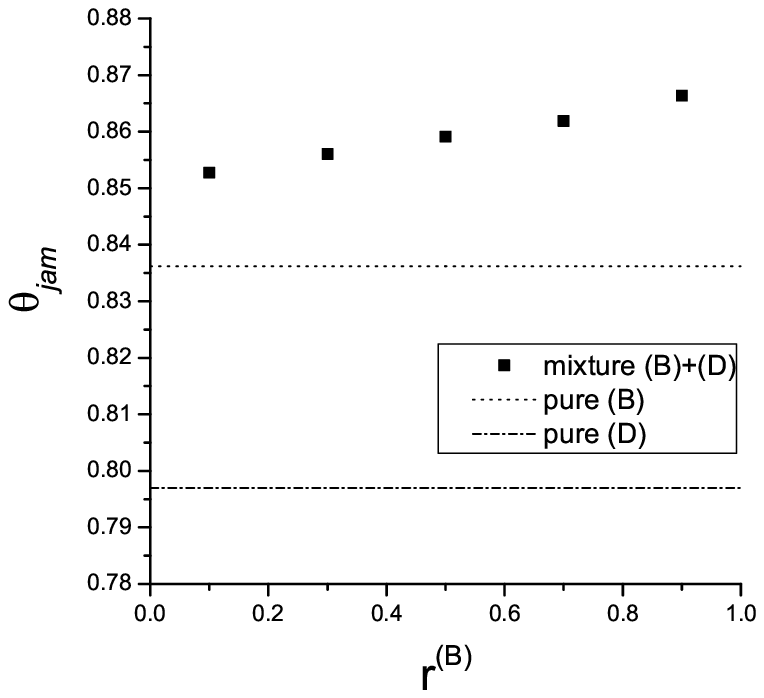}
Figure 4 a)
\end{center}
Jamming coverage for the mixture $(B)+(D)$ vs the fractional concentration
$r^{(B)}$. The dotted and the dashed lines represent the jamming coverages for
the single-component deposition of shapes $(B)$ and $(D)$, respectively.
\newpage
\begin{center}
\includegraphics[width=\columnwidth, height=\columnwidth]{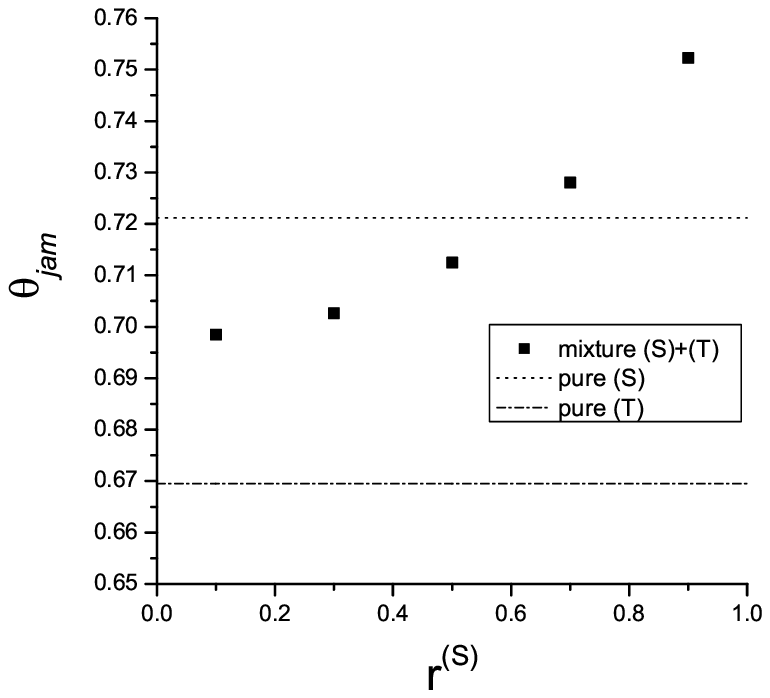}
Figure 4 b)
\end{center}
Jamming coverage for the mixture $(S)+(T)$ vs the fractional concentration
$r^{(S)}$. The dotted and the dashed lines represent the jamming coverages for
the single-component deposition of shapes $(S)$ and $(T)$, respectively.
\newpage
\begin{center}
\includegraphics[width=\columnwidth, height=\columnwidth]{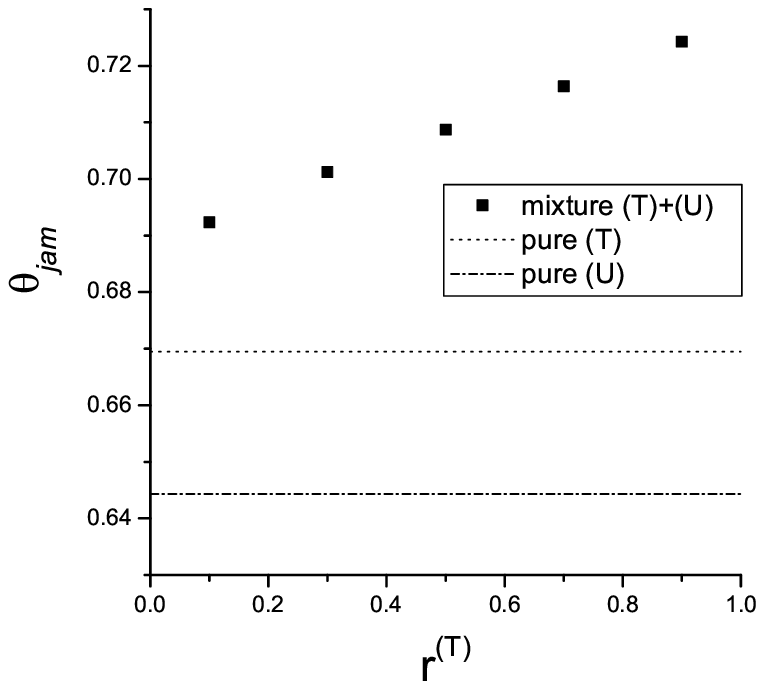}
Figure 4 c)
\end{center}
Jamming coverage for the mixture $(T)+(U)$ vs the fractional concentration
$r^{(T)}$. The dotted and the dashed lines represent the jamming coverages for
the single-component deposition of shapes $(T)$ and $(U)$, respectively.
\newpage
\begin{center}
\includegraphics[width=\columnwidth, height=\columnwidth]{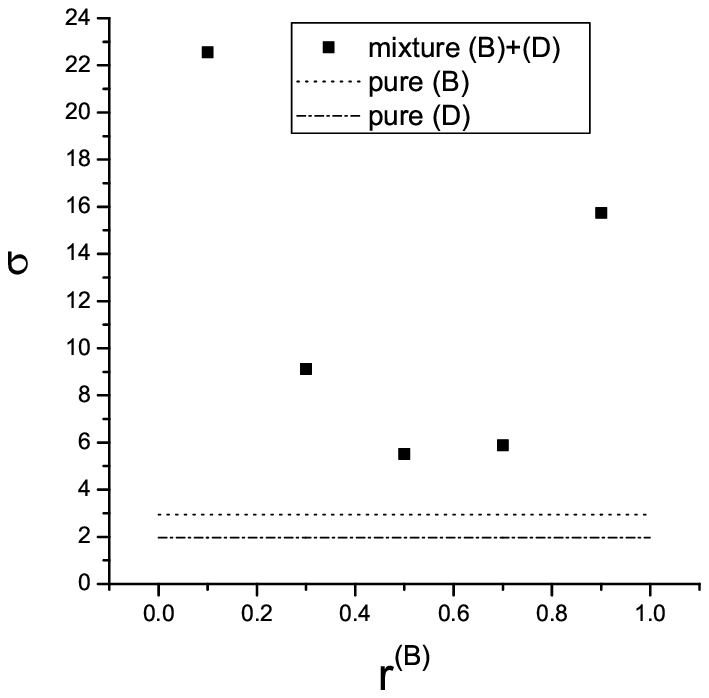}
Figure 5 a)
\end{center}
Values of the parameter $\sigma$ for the mixture $(B)+(D)$ vs the fractional
concentration $r^{(B)}$. The dotted and the dashed lines represent the values
of $\sigma$ for the single-component deposition of shapes $(B)$ and $(D)$,
respectively.
\newpage
\begin{center}
\includegraphics[width=\columnwidth, height=\columnwidth]{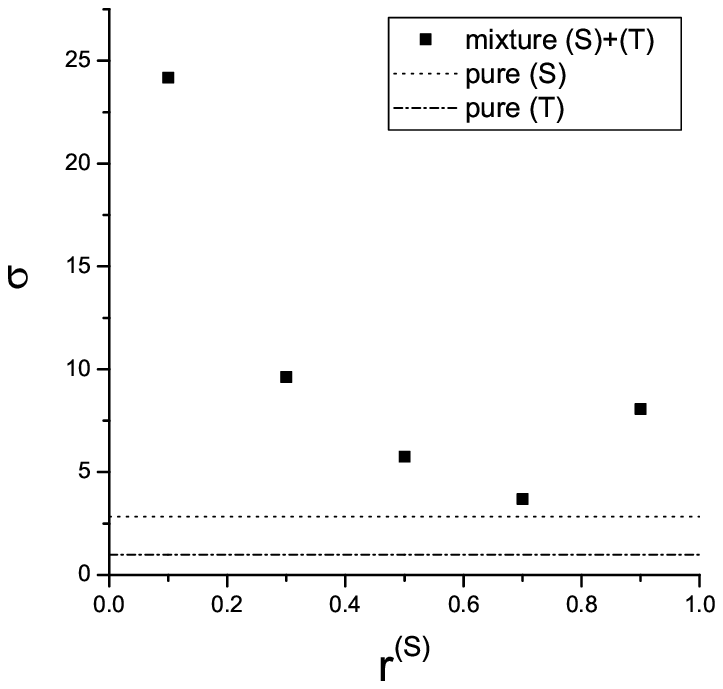}
Figure 5 b)
\end{center}
Values of the parameter $\sigma$ for the mixture $(S)+(T)$ vs the fractional
concentration $r^{(S)}$. The dotted and the dashed lines represent the values
of $\sigma$ for the single-component deposition of shapes $(S)$ and $(T)$,
respectively.
\newpage
\begin{center}
\includegraphics[width=\columnwidth, height=\columnwidth]{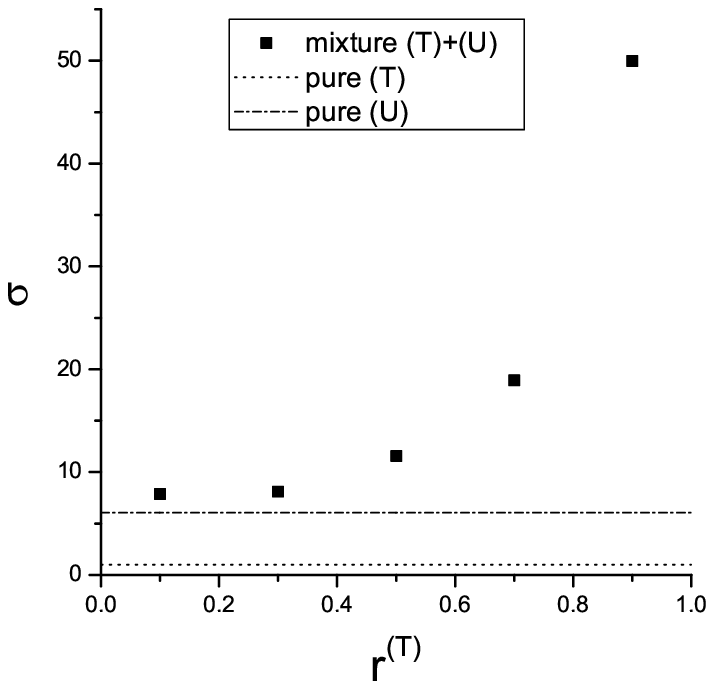}
Figure 5 c)
\end{center}
Values of the parameter $\sigma$ for the mixture $(T)+(U)$ vs the fractional
concentration $r^{(T)}$. The dotted and the dashed lines represent the values
of $\sigma$ for the single-component deposition of shapes $(T)$ and $(U)$,
respectively.
\begin{table}
\begin{tabular}{|ccc|c|c|cc|} \hline
$(x)$ & shape & $n_s^{(x)}$ & $\ell^{(x)}$ & $\theta_{jam}^{(x)}$ & $\Delta\theta$ & $\sigma$ \\ \hline
(A) &
\setlength{\unitlength}{10pt}
\begin{picture}(1,0)
\textcolor{black}{%
\put(0,0){\circle*{0.2}}
\put(0,0){\curve(0,0,1,0)}
\put(1,0){\circle*{0.2}} }
\end{picture}
& 2 & 1 & 0.9139 & 0.105 & 3.12 \\ \hline
(B) &
\setlength{\unitlength}{10pt}
\begin{picture}(2,0)
\textcolor{black}{%
\put(0,0){\circle*{0.2}}
\put(0,0){\curve(0,0,2,0)}
\put(1,0){\circle*{0.2}}
\put(2,0){\circle*{0.2}} }
\end{picture}
& 2 & & 0.8362 & 0.134 & 2.94 \\
(C) &
\setlength{\unitlength}{10pt}
\begin{picture}(1.5,1)
\textcolor{red}{%
\put(0,0){\circle*{0.2}}
\put(0,0){\curve(0,0,1,0)}
\put(1,0){\circle*{0.2}}
\put(1,0){\curve(0,0,0.5,0.866)}
\put(1.5,0.866){\circle*{0.2}} }
\end{picture}
& 1 & 2 & 0.8345 & 0.0813 & 5.78 \\
(D) &
\setlength{\unitlength}{10pt}
\begin{picture}(1,1)
\textcolor{blue}{%
\put(0,0){\circle*{0.2}}
\put(0,0){\curve(0,0,1,0)}
\put(1,0){\circle*{0.2}}
\put(1,0){\curve(0,0,-0.5,0.866)}
\put(0.5,0.866){\circle*{0.2}} }
\end{picture}
& 3 & & 0.7970 & 0.115 & 1.96 \\ \hline
(E) &
\setlength{\unitlength}{10pt}
\begin{picture}(3,0)
\textcolor{black}{%
\put(0,0){\circle*{0.2}}
\put(0,0){\curve(0,0,3,0)}
\put(1,0){\circle*{0.2}}
\put(2,0){\circle*{0.2}}
\put(3,0){\circle*{0.2}} }
\end{picture}
& 2 & & 0.7886 & 0.0944 & 3.12 \\
(F) &
\setlength{\unitlength}{10pt}
\begin{picture}(2.5,1)
\textcolor{red}{%
\put(0,0){\circle*{0.2}}
\put(0,0){\curve(0,0,2,0)}
\put(1,0){\circle*{0.2}}
\put(2,0){\circle*{0.2}}
\put(2,0){\curve(0,0,0.5,0.866)}
\put(2.5,0.866){\circle*{0.2}} }
\end{picture}
& 1 & & 0.7653 & 0.0813 & 5.74 \\
(G) &
\setlength{\unitlength}{10pt}
\begin{picture}(2,1)
\textcolor{red}{%
\put(0,0){\circle*{0.2}}
\put(0,0){\curve(0,0,2,0)}
\put(1,0){\circle*{0.2}}
\put(2,0){\circle*{0.2}}
\put(2,0){\curve(0,0,-0.5,0.866)}
\put(1.5,0.866){\circle*{0.2}} }
\end{picture}
& 1 & & 0.7739 & 0.0602 & 5.98 \\
(H) &
\setlength{\unitlength}{10pt}
\begin{picture}(2.5,1)
\textcolor{black}{%
\put(0,0){\circle*{0.2}}
\put(0,0){\curve(0,0,1,0)}
\put(1,0){\circle*{0.2}}
\put(1,0){\curve(0,0,0.5,0.866)}
\put(1.5,0.866){\circle*{0.2}}
\put(1.5,0.866){\curve(0,0,1,0)}
\put(2.5,0.866){\circle*{0.2}} }
\end{picture}
& 2 & & 0.7404 & 0.0973 & 3.03 \\
(I) &
\setlength{\unitlength}{10pt}
\begin{picture}(2,2)
\textcolor{red}{%
\put(0,0){\circle*{0.2}}
\put(0,0){\curve(0,0,1,0)}
\put(1,0){\circle*{0.2}}
\put(1,0){\curve(0,0,1,1.732 )}
\put(1.5,0.866){\circle*{0.2}}
\put(2,1.732){\circle*{0.2}} }
\end{picture}
& 1 & 3 & 0.7651 & 0.0805 & 5.75 \\
(J) &
\setlength{\unitlength}{10pt}
\begin{picture}(1.5,2)
\textcolor{red}{%
\put(0,0){\circle*{0.2}}
\put(0,0){\curve(0,0,1,0)}
\put(1,0){\circle*{0.2}}
\put(1,0){\curve(0,0,0.5,0.866)}
\put(1.5,0.866){\circle*{0.2}}
\put(1.5,0.866){\curve(0,0,-0.5,0.866)}
\put(1,1.732){\circle*{0.2}} }
\end{picture}
& 1 & & 0.7226 & 0.0573 & 5.84 \\
(K) &
\setlength{\unitlength}{10pt}
\begin{picture}(1.5,1)
\textcolor{black}{%
\put(0,0){\circle*{0.2}}
\put(0,0){\curve(0,0,1,0)}
\put(1,0){\circle*{0.2}}
\put(1,0){\curve(0,0,0.5,0.866)}
\put(1.5,0.866){\circle*{0.2}}
\put(1.5,0.866){\curve(0,0,-1,0)}
\put(0.5,0.866){\circle*{0.2}} }
\end{picture}
& 2 & & 0.7593 & 0.0590 & 2.93 \\
(L) &
\setlength{\unitlength}{10pt}
\begin{picture}(1,2)
\textcolor{red}{%
\put(0,0){\circle*{0.2}}
\put(0,0){\curve(0,0,1,0)}
\put(1,0){\circle*{0.2}}
\put(1,0){\curve(0,0,-0.5,0.866)}
\put(0.5,0.866){\circle*{0.2}}
\put(0.5,0.866){\curve(0,0,0.5,0.866)}
\put(1,1.732){\circle*{0.2}} }
\end{picture}
& 1 & & 0.7744 & 0.0578 & 6.04 \\
(M) &
\setlength{\unitlength}{10pt}
\begin{picture}(1,2)
\textcolor{red}{%
\put(0,0){\circle*{0.2}}
\put(0,0){\curve(0,0,1,0)}
\put(1,0){\circle*{0.2}}
\put(1,0){\curve(0,0,-1,1.732)}
\put(0.5,0.866){\circle*{0.2}}
\put(0,1.732){\circle*{0.2}} }
\end{picture}
& 1 & & 0.7742 & 0.0608 & 5.96 \\ \hline
(N) &
\setlength{\unitlength}{10pt}
\begin{picture}(4,0)
\textcolor{black}{%
\put(0,0){\circle*{0.2}}
\put(0,0){\curve(0,0,4,0)}
\put(1,0){\circle*{0.2}}
\put(2,0){\circle*{0.2}}
\put(3,0){\circle*{0.2}}
\put(4,0){\circle*{0.2}} }
\end{picture}
& 2 & 4 & 0.7587 & 0.115 & 2.87 \\ \hline
(O) &
\setlength{\unitlength}{10pt}
\begin{picture}(5,0)
\textcolor{black}{%
\put(0,0){\circle*{0.2}}
\put(0,0){\curve(0,0,5,0)}
\put(1,0){\circle*{0.2}}
\put(2,0){\circle*{0.2}}
\put(3,0){\circle*{0.2}}
\put(4,0){\circle*{0.2}}
\put(5,0){\circle*{0.2}} }
\end{picture}
& 2 & & 0.7372 & 0.0837 & 3.02 \\
(P) &
\setlength{\unitlength}{10pt}
\begin{picture}(2,2)
\textcolor{blue}{%
\put(0,0){\circle*{0.2}}
\put(0,0){\curve(0,0,2,0)}
\put(1,0){\circle*{0.2}}
\put(2,0){\circle*{0.2}}
\put(2,0){\curve(0,0,-0.5,0.866)}
\put(1.5,0.866){\circle*{0.2}}
\put(1.5,0.866){\curve(0,0,-1,0)}
\put(0.5,0.866){\circle*{0.2}}
\put(0.5,0.866){\curve(0,0,0.5,0.866)}
\put(1,1.732){\circle*{0.2}} }
\end{picture}
& 3 & & 0.7210 & 0.0944 & 1.99 \\
(Q) &
\setlength{\unitlength}{10pt}
\begin{picture}(2,2)
\textcolor{green}{%
\put(0,0){\circle*{0.2}}
\put(0,0){\curve(0,0,1,0)}
\put(1,0){\circle*{0.2}}
\put(1,0){\curve(0,0,0.5,0.866)}
\put(1.5,0.866){\circle*{0.2}}
\put(1.5,0.866){\curve(0,0,-0.5,0.866)}
\put(1,1.732){\circle*{0.2}}
\put(1,1.732){\curve(0,0,-1,0)}
\put(0,1.732){\circle*{0.2}}
\put(0,1.732){\curve(0,0,-0.5,-0.866)}
\put(-0.5,0.866){\circle*{0.2}} }
\end{picture}
& 6 & \raisebox{2.5ex}[0pt]{5} & 0.5740 & 0.0707 & 0.986 \\
(R) &
\setlength{\unitlength}{10pt}
\begin{picture}(4,2)
\textcolor{red}{%
\put(0,0){\circle*{0.2}}
\put(0,0){\curve(0,0,3,0)}
\put(1,0){\circle*{0.2}}
\put(2,0){\circle*{0.2}}
\put(3,0){\circle*{0.2}}
\put(3,0){\curve(0,0,1,1.732)}
\put(3.5,0.866){\circle*{0.2}}
\put(4,1.732){\circle*{0.2}} }
\end{picture}
& 1 & & 0.6758 & 0.0829 & 5.81 \\ \hline
(S) &
\setlength{\unitlength}{10pt}
\begin{picture}(6,0)
\textcolor{black}{%
\put(0,0){\circle*{0.2}}
\put(0,0){\curve(0,0,6,0)}
\put(1,0){\circle*{0.2}}
\put(2,0){\circle*{0.2}}
\put(3,0){\circle*{0.2}}
\put(4,0){\circle*{0.2}}
\put(5,0){\circle*{0.2}}
\put(6,0){\circle*{0.2}} }
\end{picture}
& 2 & & 0.7212 & 0.0954 & 2.84 \\
(T) &
\setlength{\unitlength}{10pt}
\begin{picture}(2,2)
\textcolor{green}{%
\put(0,0){\circle*{0.2}}
\put(0,0){\curve(0,0,1,0)}
\put(1,0){\circle*{0.2}}
\put(1,0){\curve(0,0,0.5,0.866)}
\put(1.5,0.866){\circle*{0.2}}
\put(1.5,0.866){\curve(0,0,-2,0)}
\put(0.5,0.866){\circle*{0.2}}
\put(-0.5,0.866){\circle*{0.2}}
\put(-0.5,0.866){\curve(0,0,0.5,0.866)}
\put(0,1.732){\circle*{0.2}}
\put(0,1.732){\curve(0,0,1,0)}
\put(1,1.732){\circle*{0.2}} }
\end{picture}
& 6 & & 0.6695 & 0.0773 & 0.994 \\
(U) &
\setlength{\unitlength}{10pt}
\begin{picture}(5,2)
\textcolor{red}{%
\put(0,0){\circle*{0.2}}
\put(0,0){\curve(0,0,4,0)}
\put(1,0){\circle*{0.2}}
\put(2,0){\circle*{0.2}}
\put(3,0){\circle*{0.2}}
\put(4,0){\circle*{0.2}}
\put(4,0){\curve(0,0,1,1.732)}
\put(4.5,0.866){\circle*{0.2}}
\put(5,1.732){\circle*{0.2}} }
\end{picture}
& 1 & \raisebox{2.5ex}[0pt]{6} & 0.6443 & 0.0721 & 6.07 \\ \hline
\end{tabular}
\caption{Parameters $\Delta\theta$ and $\sigma$ determined using
Eq.~\eqref{eq:rhoIRSA} for various shapes $(x)$ of length $\ell$ on a
triangular lattice. The colors (online only) are associated with the different
order $n^{(x)}_s$ of symmetry axis.  The typical statistical errors are
estimated to the last given digits.}
\label{tab:pure_objects}
\end{table}
\begin{table}
\begin{tabular}{|ccc|c|c|cc|c|c|} \hline
$(x) + (y)$ & shapes & $n_s^{(x)} + n_s^{(y)}$ & $\ell^{(x)} + \ell^{(y)}$ & $\theta_{jam}^{(x)+(y)}$ & $\theta_{\infty}^{(x)}$ & $\theta_{\infty}^{(y)}$ & $\Delta\theta$ & $\sigma$ \\ \hline
$(A)$ + $(B)$ &
\setlength{\unitlength}{10pt}
\begin{picture}(1,0)
\textcolor{black}{%
\put(0,0){\circle*{0.2}}
\put(0,0){\curve(0,0,1,0)}
\put(1,0){\circle*{0.2}} }
\end{picture}
+
\setlength{\unitlength}{10pt}
\begin{picture}(2,0)
\textcolor{black}{%
\put(0,0){\circle*{0.2}}
\put(0,0){\curve(0,0,2,0)}
\put(1,0){\circle*{0.2}}
\put(2,0){\circle*{0.2}} }
\end{picture}
& 2 + 2 & 1 + 2 & 0.9202 & 0.5401 & 0.3801 & 0.125 & 5.52 \\
$(A)$ + $(E)$ &
\setlength{\unitlength}{10pt}
\begin{picture}(1,0)
\textcolor{black}{%
\put(0,0){\circle*{0.2}}
\put(0,0){\curve(0,0,1,0)}
\put(1,0){\circle*{0.2}} }
\end{picture}
+
\setlength{\unitlength}{10pt}
\begin{picture}(3,0)
\textcolor{black}{%
\put(0,0){\circle*{0.2}}
\put(0,0){\curve(0,0,3,0)}
\put(1,0){\circle*{0.2}}
\put(2,0){\circle*{0.2}}
\put(3,0){\circle*{0.2}} }
\end{picture}
& 2 + 2 & 1 + 3 & 0.9191 & 0.5833 & 0.3358 & 0.129 & 5.66 \\
$(A)$ + $(N)$ &
\setlength{\unitlength}{10pt}
\begin{picture}(1,0)
\textcolor{black}{%
\put(0,0){\circle*{0.2}}
\put(0,0){\curve(0,0,1,0)}
\put(1,0){\circle*{0.2}} }
\end{picture}
+
\setlength{\unitlength}{10pt}
\begin{picture}(4,0)
\textcolor{black}{%
\put(0,0){\circle*{0.2}}
\put(0,0){\curve(0,0,4,0)}
\put(1,0){\circle*{0.2}}
\put(2,0){\circle*{0.2}}
\put(3,0){\circle*{0.2}}
\put(4,0){\circle*{0.2}} }
\end{picture}
& 2 + 2 & 1 + 4 & 0.9196 & 0.6144 & 0.3052 & 0.131 & 5.77 \\
$(A)$ + $(O)$ &
\setlength{\unitlength}{10pt}
\begin{picture}(1,0)
\textcolor{black}{%
\put(0,0){\circle*{0.2}}
\put(0,0){\curve(0,0,1,0)}
\put(1,0){\circle*{0.2}} }
\end{picture}
+
\setlength{\unitlength}{10pt}
\begin{picture}(5,0)
\textcolor{black}{%
\put(0,0){\circle*{0.2}}
\put(0,0){\curve(0,0,5,0)}
\put(1,0){\circle*{0.2}}
\put(2,0){\circle*{0.2}}
\put(3,0){\circle*{0.2}}
\put(4,0){\circle*{0.2}}
\put(5,0){\circle*{0.2}} }
\end{picture}
& 2 + 2 & 1 + 5 & 0.9195 & 0.6143 & 0.3052 & 0.135 & 5.73 \\
$(A)$ + $(S)$ &
\setlength{\unitlength}{10pt}
\begin{picture}(1,0)
\textcolor{black}{%
\put(0,0){\circle*{0.2}}
\put(0,0){\curve(0,0,1,0)}
\put(1,0){\circle*{0.2}} }
\end{picture}
+
\setlength{\unitlength}{10pt}
\begin{picture}(6,0)
\textcolor{black}{%
\put(0,0){\circle*{0.2}}
\put(0,0){\curve(0,0,6,0)}
\put(1,0){\circle*{0.2}}
\put(2,0){\circle*{0.2}}
\put(3,0){\circle*{0.2}}
\put(4,0){\circle*{0.2}}
\put(5,0){\circle*{0.2}}
\put(6,0){\circle*{0.2}} }
\end{picture}
& 2 + 2 & 1 + 6 & 0.9198 & 0.6553 & 0.2645 & 0.157 & 5.76 \\ \hline
$(B)$ + $(C)$ &
\setlength{\unitlength}{10pt}
\begin{picture}(2,0)
\textcolor{black}{%
\put(0,0){\circle*{0.2}}
\put(0,0){\curve(0,0,2,0)}
\put(1,0){\circle*{0.2}}
\put(2,0){\circle*{0.2}} }
\end{picture}
+
\setlength{\unitlength}{10pt}
\begin{picture}(1.5,1)
\textcolor{red}{%
\put(0,0){\circle*{0.2}}
\put(0,0){\curve(0,0,1,0)}
\put(1,0){\circle*{0.2}}
\put(1,0){\curve(0,0,0.5,0.866)}
\put(1.5,0.866){\circle*{0.2}} }
\end{picture}
& 2 + 1 & 2 + 2 & 0.8526 & 0.4191 & 0.4335 & 0.0365 & 11.42 \\
$(B)$ + $(D)$ &
\setlength{\unitlength}{10pt}
\begin{picture}(2,0)
\textcolor{black}{%
\put(0,0){\circle*{0.2}}
\put(0,0){\curve(0,0,2,0)}
\put(1,0){\circle*{0.2}}
\put(2,0){\circle*{0.2}} }
\end{picture}
+
\setlength{\unitlength}{10pt}
\begin{picture}(1,1)
\textcolor{blue}{%
\put(0,0){\circle*{0.2}}
\put(0,0){\curve(0,0,1,0)}
\put(1,0){\circle*{0.2}}
\put(1,0){\curve(0,0,-0.5,0.866)}
\put(0.5,0.866){\circle*{0.2}} }
\end{picture}
& 2 + 3 & 2 + 2 & 0.8591 & 0.4330 & 0.4261 & 0.0781 & 5.51 \\
$(C)$ + $(D)$ &
\setlength{\unitlength}{10pt}
\begin{picture}(1.5,1)
\textcolor{red}{%
\put(0,0){\circle*{0.2}}
\put(0,0){\curve(0,0,1,0)}
\put(1,0){\circle*{0.2}}
\put(1,0){\curve(0,0,0.5,0.866)}
\put(1.5,0.866){\circle*{0.2}} }
\end{picture}
+
\setlength{\unitlength}{10pt}
\begin{picture}(1,1)
\textcolor{blue}{%
\put(0,0){\circle*{0.2}}
\put(0,0){\curve(0,0,1,0)}
\put(1,0){\circle*{0.2}}
\put(1,0){\curve(0,0,-0.5,0.866)}
\put(0.5,0.866){\circle*{0.2}} }
\end{picture}
& 1 + 3 & 2 + 2 & 0.8624 & 0.4406 & 0.4218 & 0.0493 & 11.62 \\
$(D)$ + $(P)$ &
\setlength{\unitlength}{10pt}
\begin{picture}(1,1)
\textcolor{blue}{%
\put(0,0){\circle*{0.2}}
\put(0,0){\curve(0,0,1,0)}
\put(1,0){\circle*{0.2}}
\put(1,0){\curve(0,0,-0.5,0.866)}
\put(0.5,0.866){\circle*{0.2}} }
\end{picture}
+
\setlength{\unitlength}{10pt}
\begin{picture}(2,2)
\textcolor{blue}{%
\put(0,0){\circle*{0.2}}
\put(0,0){\curve(0,0,2,0)}
\put(1,0){\circle*{0.2}}
\put(2,0){\circle*{0.2}}
\put(2,0){\curve(0,0,-0.5,0.866)}
\put(1.5,0.866){\circle*{0.2}}
\put(1.5,0.866){\curve(0,0,-1,0)}
\put(0.5,0.866){\circle*{0.2}}
\put(0.5,0.866){\curve(0,0,0.5,0.866)}
\put(1,1.732){\circle*{0.2}} }
\end{picture}
& 3 + 3 & 2 + 5 & 0.8211 & 0.4946 & 0.3265 & 0.0983 & 3.95 \\ \hline
$(E)$ + $(F)$ &
\setlength{\unitlength}{10pt}
\begin{picture}(3,0)
\textcolor{black}{%
\put(0,0){\circle*{0.2}}
\put(0,0){\curve(0,0,3,0)}
\put(1,0){\circle*{0.2}}
\put(2,0){\circle*{0.2}}
\put(3,0){\circle*{0.2}} }
\end{picture}
+
\setlength{\unitlength}{10pt}
\begin{picture}(2.5,1)
\textcolor{red}{%
\put(0,0){\circle*{0.2}}
\put(0,0){\curve(0,0,2,0)}
\put(1,0){\circle*{0.2}}
\put(2,0){\circle*{0.2}}
\put(2,0){\curve(0,0,0.5,0.866)}
\put(2.5,0.866){\circle*{0.2}} }
\end{picture}
& 2 + 1 & 3 + 3 & 0.7876 & 0.3903 & 0.3973 & 0.0354 & 11.84 \\
$(E)$ + $(K)$ &
\setlength{\unitlength}{10pt}
\begin{picture}(3,0)
\textcolor{black}{%
\put(0,0){\circle*{0.2}}
\put(0,0){\curve(0,0,3,0)}
\put(1,0){\circle*{0.2}}
\put(2,0){\circle*{0.2}}
\put(3,0){\circle*{0.2}} }
\end{picture}
+
\setlength{\unitlength}{10pt}
\begin{picture}(1.5,1)
\textcolor{black}{%
\put(0,0){\circle*{0.2}}
\put(0,0){\curve(0,0,1,0)}
\put(1,0){\circle*{0.2}}
\put(1,0){\curve(0,0,0.5,0.866)}
\put(1.5,0.866){\circle*{0.2}}
\put(1.5,0.866){\curve(0,0,-1,0)}
\put(0.5,0.866){\circle*{0.2}} }
\end{picture}
& 2 + 2 & 3 + 3 & 0.8109 & 0.3917 & 0.4192 & 0.0620 & 5.87 \\
$(F)$ + $(K)$ &
\setlength{\unitlength}{10pt}
\begin{picture}(2.5,1)
\textcolor{red}{%
\put(0,0){\circle*{0.2}}
\put(0,0){\curve(0,0,2,0)}
\put(1,0){\circle*{0.2}}
\put(2,0){\circle*{0.2}}
\put(2,0){\curve(0,0,0.5,0.866)}
\put(2.5,0.866){\circle*{0.2}} }
\end{picture}
+
\setlength{\unitlength}{10pt}
\begin{picture}(1.5,1)
\textcolor{black}{%
\put(0,0){\circle*{0.2}}
\put(0,0){\curve(0,0,1,0)}
\put(1,0){\circle*{0.2}}
\put(1,0){\curve(0,0,0.5,0.866)}
\put(1.5,0.866){\circle*{0.2}}
\put(1.5,0.866){\curve(0,0,-1,0)}
\put(0.5,0.866){\circle*{0.2}} }
\end{picture}
& 1 + 2 & 3 + 3 & 0.8140 & 0.3960 & 0.4180 & 0.0523 & 11.55 \\ \hline
$(O)$ + $(P)$ &
\setlength{\unitlength}{10pt}
\begin{picture}(5,0)
\textcolor{black}{%
\put(0,0){\circle*{0.2}}
\put(0,0){\curve(0,0,5,0)}
\put(1,0){\circle*{0.2}}
\put(2,0){\circle*{0.2}}
\put(3,0){\circle*{0.2}}
\put(4,0){\circle*{0.2}}
\put(5,0){\circle*{0.2}} }
\end{picture}
+
\setlength{\unitlength}{10pt}
\begin{picture}(2,2)
\textcolor{blue}{%
\put(0,0){\circle*{0.2}}
\put(0,0){\curve(0,0,2,0)}
\put(1,0){\circle*{0.2}}
\put(2,0){\circle*{0.2}}
\put(2,0){\curve(0,0,-0.5,0.866)}
\put(1.5,0.866){\circle*{0.2}}
\put(1.5,0.866){\curve(0,0,-1,0)}
\put(0.5,0.866){\circle*{0.2}}
\put(0.5,0.866){\curve(0,0,0.5,0.866)}
\put(1,1.732){\circle*{0.2}} }
\end{picture}
& 2 + 3 & 5 + 5 & 0.7647 & 0.3415 & 0.4232 & 0.0639 & 5.41 \\
$(O)$ + $(Q)$ &
\setlength{\unitlength}{10pt}
\begin{picture}(5,0)
\textcolor{black}{%
\put(0,0){\circle*{0.2}}
\put(0,0){\curve(0,0,5,0)}
\put(1,0){\circle*{0.2}}
\put(2,0){\circle*{0.2}}
\put(3,0){\circle*{0.2}}
\put(4,0){\circle*{0.2}}
\put(5,0){\circle*{0.2}} }
\end{picture}
+
\setlength{\unitlength}{10pt}
\begin{picture}(2,2)
\textcolor{green}{%
\put(0,0){\circle*{0.2}}
\put(0,0){\curve(0,0,1,0)}
\put(1,0){\circle*{0.2}}
\put(1,0){\curve(0,0,0.5,0.866)}
\put(1.5,0.866){\circle*{0.2}}
\put(1.5,0.866){\curve(0,0,-0.5,0.866)}
\put(1,1.732){\circle*{0.2}}
\put(1,1.732){\curve(0,0,-1,0)}
\put(0,1.732){\circle*{0.2}}
\put(0,1.732){\curve(0,0,-0.5,-0.866)}
\put(-0.5,0.866){\circle*{0.2}} }
\end{picture}
& 2 + 6 & 5 + 5 & 0.6791 & 0.3608 & 0.3183 & 0.0534 & 5.54 \\
$(O)$ + $(R)$ &
\setlength{\unitlength}{10pt}
\begin{picture}(5,0)
\textcolor{black}{%
\put(0,0){\circle*{0.2}}
\put(0,0){\curve(0,0,5,0)}
\put(1,0){\circle*{0.2}}
\put(2,0){\circle*{0.2}}
\put(3,0){\circle*{0.2}}
\put(4,0){\circle*{0.2}}
\put(5,0){\circle*{0.2}} }
\end{picture}
+
\setlength{\unitlength}{10pt}
\begin{picture}(4,2)
\textcolor{red}{%
\put(0,0){\circle*{0.2}}
\put(0,0){\curve(0,0,3,0)}
\put(1,0){\circle*{0.2}}
\put(2,0){\circle*{0.2}}
\put(3,0){\circle*{0.2}}
\put(3,0){\curve(0,0,1,1.732)}
\put(3.5,0.866){\circle*{0.2}}
\put(4,1.732){\circle*{0.2}} }
\end{picture}
& 2 + 1 & 5 + 5 & 0.7086 & 0.3612 & 0.3474 & 0.0460 & 11.21 \\
$(P)$ + $(Q)$ &
\setlength{\unitlength}{10pt}
\begin{picture}(2,2)
\textcolor{blue}{%
\put(0,0){\circle*{0.2}}
\put(0,0){\curve(0,0,2,0)}
\put(1,0){\circle*{0.2}}
\put(2,0){\circle*{0.2}}
\put(2,0){\curve(0,0,-0.5,0.866)}
\put(1.5,0.866){\circle*{0.2}}
\put(1.5,0.866){\curve(0,0,-1,0)}
\put(0.5,0.866){\circle*{0.2}}
\put(0.5,0.866){\curve(0,0,0.5,0.866)}
\put(1,1.732){\circle*{0.2}} }
\end{picture}
+
\setlength{\unitlength}{10pt}
\begin{picture}(2,2)
\textcolor{green}{%
\put(0,0){\circle*{0.2}}
\put(0,0){\curve(0,0,1,0)}
\put(1,0){\circle*{0.2}}
\put(1,0){\curve(0,0,0.5,0.866)}
\put(1.5,0.866){\circle*{0.2}}
\put(1.5,0.866){\curve(0,0,-0.5,0.866)}
\put(1,1.732){\circle*{0.2}}
\put(1,1.732){\curve(0,0,-1,0)}
\put(0,1.732){\circle*{0.2}}
\put(0,1.732){\curve(0,0,-0.5,-0.866)}
\put(-0.5,0.866){\circle*{0.2}} }
\end{picture}
& 3 + 6 & 5 + 5 & 0.6833 & 0.3998 & 0.2835 & 0.0686 & 3.98 \\
$(P)$ + $(R)$ &
\setlength{\unitlength}{10pt}
\begin{picture}(2,2)
\textcolor{blue}{%
\put(0,0){\circle*{0.2}}
\put(0,0){\curve(0,0,2,0)}
\put(1,0){\circle*{0.2}}
\put(2,0){\circle*{0.2}}
\put(2,0){\curve(0,0,-0.5,0.866)}
\put(1.5,0.866){\circle*{0.2}}
\put(1.5,0.866){\curve(0,0,-1,0)}
\put(0.5,0.866){\circle*{0.2}}
\put(0.5,0.866){\curve(0,0,0.5,0.866)}
\put(1,1.732){\circle*{0.2}} }
\end{picture}
+
\setlength{\unitlength}{10pt}
\begin{picture}(4,2)
\textcolor{red}{%
\put(0,0){\circle*{0.2}}
\put(0,0){\curve(0,0,3,0)}
\put(1,0){\circle*{0.2}}
\put(2,0){\circle*{0.2}}
\put(3,0){\circle*{0.2}}
\put(3,0){\curve(0,0,1,1.732)}
\put(3.5,0.866){\circle*{0.2}}
\put(4,1.732){\circle*{0.2}} }
\end{picture}
& 3 + 1 & 5 + 5 & 0.7485 & 0.4156 & 0.3329 & 0.0420 & 11.74 \\
$(Q)$ + $(R)$ &
\setlength{\unitlength}{10pt}
\begin{picture}(2,2)
\textcolor{green}{%
\put(0,0){\circle*{0.2}}
\put(0,0){\curve(0,0,1,0)}
\put(1,0){\circle*{0.2}}
\put(1,0){\curve(0,0,0.5,0.866)}
\put(1.5,0.866){\circle*{0.2}}
\put(1.5,0.866){\curve(0,0,-0.5,0.866)}
\put(1,1.732){\circle*{0.2}}
\put(1,1.732){\curve(0,0,-1,0)}
\put(0,1.732){\circle*{0.2}}
\put(0,1.732){\curve(0,0,-0.5,-0.866)}
\put(-0.5,0.866){\circle*{0.2}} }
\end{picture}
+
\setlength{\unitlength}{10pt}
\begin{picture}(4,2)
\textcolor{red}{%
\put(0,0){\circle*{0.2}}
\put(0,0){\curve(0,0,3,0)}
\put(1,0){\circle*{0.2}}
\put(2,0){\circle*{0.2}}
\put(3,0){\circle*{0.2}}
\put(3,0){\curve(0,0,1,1.732)}
\put(3.5,0.866){\circle*{0.2}}
\put(4,1.732){\circle*{0.2}} }
\end{picture}
& 6 + 1 & 5 + 5 & 0.6822 & 0.3227 & 0.3595 & 0.0429 & 12.07 \\
$(Q)$ + $(T)$ &
\setlength{\unitlength}{10pt}
\begin{picture}(2,2)
\textcolor{green}{%
\put(0,0){\circle*{0.2}}
\put(0,0){\curve(0,0,1,0)}
\put(1,0){\circle*{0.2}}
\put(1,0){\curve(0,0,0.5,0.866)}
\put(1.5,0.866){\circle*{0.2}}
\put(1.5,0.866){\curve(0,0,-0.5,0.866)}
\put(1,1.732){\circle*{0.2}}
\put(1,1.732){\curve(0,0,-1,0)}
\put(0,1.732){\circle*{0.2}}
\put(0,1.732){\curve(0,0,-0.5,-0.866)}
\put(-0.5,0.866){\circle*{0.2}} }
\end{picture}
+
\setlength{\unitlength}{10pt}
\begin{picture}(2,2)
\textcolor{green}{%
\put(0,0){\circle*{0.2}}
\put(0,0){\curve(0,0,1,0)}
\put(1,0){\circle*{0.2}}
\put(1,0){\curve(0,0,0.5,0.866)}
\put(1.5,0.866){\circle*{0.2}}
\put(1.5,0.866){\curve(0,0,-2,0)}
\put(0.5,0.866){\circle*{0.2}}
\put(-0.5,0.866){\circle*{0.2}}
\put(-0.5,0.866){\curve(0,0,0.5,0.866)}
\put(0,1.732){\circle*{0.2}}
\put(0,1.732){\curve(0,0,1,0)}
\put(1,1.732){\circle*{0.2}} }
\end{picture}
& 6 + 6 & 5 + 6 & 0.6222 & 0.2873 & 0.3349 & 0.0758 & 0.985 \\ \hline
$(S)$ + $(T)$ &
\setlength{\unitlength}{10pt}
\begin{picture}(6,0)
\textcolor{black}{%
\put(0,0){\circle*{0.2}}
\put(0,0){\curve(0,0,6,0)}
\put(1,0){\circle*{0.2}}
\put(2,0){\circle*{0.2}}
\put(3,0){\circle*{0.2}}
\put(4,0){\circle*{0.2}}
\put(5,0){\circle*{0.2}}
\put(6,0){\circle*{0.2}} }
\end{picture}
+
\setlength{\unitlength}{10pt}
\begin{picture}(2,2)
\textcolor{green}{%
\put(0,0){\circle*{0.2}}
\put(0,0){\curve(0,0,1,0)}
\put(1,0){\circle*{0.2}}
\put(1,0){\curve(0,0,0.5,0.866)}
\put(1.5,0.866){\circle*{0.2}}
\put(1.5,0.866){\curve(0,0,-2,0)}
\put(0.5,0.866){\circle*{0.2}}
\put(-0.5,0.866){\circle*{0.2}}
\put(-0.5,0.866){\curve(0,0,0.5,0.866)}
\put(0,1.732){\circle*{0.2}}
\put(0,1.732){\curve(0,0,1,0)}
\put(1,1.732){\circle*{0.2}} }
\end{picture}
& 2 + 6 & 6 + 6 & 0.7125 & 0.3218 & 0.3907 & 0.0344 & 5.76 \\
$(S)$ + $(U)$ &
\setlength{\unitlength}{10pt}
\begin{picture}(6,0)
\textcolor{black}{%
\put(0,0){\circle*{0.2}}
\put(0,0){\curve(0,0,6,0)}
\put(1,0){\circle*{0.2}}
\put(2,0){\circle*{0.2}}
\put(3,0){\circle*{0.2}}
\put(4,0){\circle*{0.2}}
\put(5,0){\circle*{0.2}}
\put(6,0){\circle*{0.2}} }
\end{picture}
+
\setlength{\unitlength}{10pt}
\begin{picture}(5,2)
\textcolor{red}{%
\put(0,0){\circle*{0.2}}
\put(0,0){\curve(0,0,4,0)}
\put(1,0){\circle*{0.2}}
\put(2,0){\circle*{0.2}}
\put(3,0){\circle*{0.2}}
\put(4,0){\circle*{0.2}}
\put(4,0){\curve(0,0,1,1.732)}
\put(4.5,0.866){\circle*{0.2}}
\put(5,1.732){\circle*{0.2}} }
\end{picture}
& 2 + 1 & 6 + 6 & 0.6833 & 0.3533 & 0.3300 & 0.0437 & 11.22 \\
$(T)$ + $(U)$ &
\setlength{\unitlength}{10pt}
\begin{picture}(2,2)
\textcolor{green}{%
\put(0,0){\circle*{0.2}}
\put(0,0){\curve(0,0,1,0)}
\put(1,0){\circle*{0.2}}
\put(1,0){\curve(0,0,0.5,0.866)}
\put(1.5,0.866){\circle*{0.2}}
\put(1.5,0.866){\curve(0,0,-2,0)}
\put(0.5,0.866){\circle*{0.2}}
\put(-0.5,0.866){\circle*{0.2}}
\put(-0.5,0.866){\curve(0,0,0.5,0.866)}
\put(0,1.732){\circle*{0.2}}
\put(0,1.732){\curve(0,0,1,0)}
\put(1,1.732){\circle*{0.2}} }
\end{picture}
+
\setlength{\unitlength}{10pt}
\begin{picture}(5,2)
\textcolor{red}{%
\put(0,0){\circle*{0.2}}
\put(0,0){\curve(0,0,4,0)}
\put(1,0){\circle*{0.2}}
\put(2,0){\circle*{0.2}}
\put(3,0){\circle*{0.2}}
\put(4,0){\circle*{0.2}}
\put(4,0){\curve(0,0,1,1.732)}
\put(4.5,0.866){\circle*{0.2}}
\put(5,1.732){\circle*{0.2}} }
\end{picture}
& 6 + 1 & 6 + 6 & 0.7087 & 0.3955 & 0.3132 & 0.0450 & 11.57 \\ \hline
\end{tabular}
\caption{Coverage fraction $\theta_{jam}^{(x)+(y)}$ for various binary mixtures
$(x)+(y)$  of shapes $A$ -- $V$ from Table I.  The colors (online only) are
associated with the different order $n_s^{(x)}$ of symmetry axis. The typical
statistical errors are estimated to the last given digits.}
\label{tab:mixtures}
\end{table}
\end{document}